\documentclass[conference]{IEEEtran}
\usepackage{cite}
\usepackage{amsmath,amssymb,amsfonts}
\usepackage{algorithmic}
\usepackage{graphicx}
\usepackage{textcomp}
\usepackage{xcolor}
\def\BibTeX{{\rm B\kern-.05em{\sc i\kern-.025em b}\kern-.08em
    T\kern-.1667em\lower.7ex\hbox{E}\kern-.125emX}}
\usepackage{listings}

\begin{document}

\title{Programming of Cellular Automata in C and C++\\
}

\author{\IEEEauthorblockN{Patrik Christen}
\IEEEauthorblockA{\textit{Institute for Information Systems} \\
\textit{FHNW}\\
Olten, Switzerland \\
patrik.christen@fhnw.ch \\
and \\
\textit{Chair for Philosophy} \\
\textit{ETH Zurich}\\
Zurich, Switzerland \\
patrik.christen@gess.ethz.ch}
}

\maketitle

\begin{abstract}
This study explores running times of different ways to program cellular automata in C and C++, i.e. looping through arrays by different means, the effect of structures and objects, and the choice of data structure (array versus vector in C++) and compiler (GNU gcc versus Apple clang). Using arrays instead of vectors, using pointers instead of array indices, using C instead of C++, and using structures and objects instead of primitive data types has little to negligible effects on the running time. Also, the choice of compiler has only a minor effect, except for simple update functions, in which case clang seems to find better ways to optimise it, especially for index-based access. If one is interested in multi-state cellular automata, structures and objects can be used without loss of performance in C and C++, respectively. The study supports the recommendation from practice to use vector in C++ and adds to it the use of index-based access for cellular automata. Future studies might investigate Apple's Metal shader or compiler optimisation, especially with respect to the update function.
\end{abstract}


\section{Introduction}
It is known since a long time that simple programs such as elementary cellular automata can create complex patterns \cite{Wolfram.Nature.1984,Wolfram.NKS.2002}. The astonishing thing about this is that the computer program of a cellular automaton creating a simple, periodic, random, or complex pattern does not differ in complexity. Only the local update rule or function differs, specifically the values in a Boolean array with eight elements defining the rule. It is a surprising and important result because it shows that software does not have to be programmed in a complex way to create complex patterns or to solve complex problems. In general, and with cellular automata in particular, it should therefore be possible to create simple and small programs capable of solving complex problems in science and industry. This should even be possible to run on mobile and desktop computers, which is interesting for edge computing. Making efficient use of the ever more performant computer hardware for desktop computing, large-scale simulations can be envisioned as well.

So far, most of the software development does not make use of any of this. On the contrary, software gets more and more complex not because of the complexity of the problem to be solved but because of its ever-increasing size and number of dependencies \cite{Hubert.IEEESpectrum.2024} as well as the way we program it \cite{Wirth.Computer.1995,Wirth.IEEEAHC.2008}. Already in 1995, Niklaus Wirth pointed out in \cite{Wirth.Computer.1995} that "Software expands to fill the available memory." and "Software is getting slower more rapidly than hardware becomes faster.". Some years later he also pointed out that we have lost sight of the need for careful craftsmanship in programming, even at universities \cite{Wirth.IEEEAHC.2008}. Almost twenty years later, most companies seem to focus more on making feature-rich and shiny software than on performance, quality, or tackling the complexity problem in general. Furthermore, high-level languages and easy-to-use frameworks with uncountable packages are used, which leads to patch-work code and to numerous dependencies \cite{Hubert.IEEESpectrum.2024}. The situation has become even worse with the recent advent of code generation with chatbots and the integration of artificial intelligence components into virtually every software.

In the present paper I want to rescue some of the programming craftsmanship with respect to cellular automata although the results are applicable to other computer programs as well. Specifically, I am investigating different ways to use arrays to program cellular automata in C and C++ and compare their running time for one update iteration. Applications of cellular automata are often consisting of a large number of elements since this is a typical property of many complex systems. In addition, most applications require experimenting with different cellular automaton configurations. It is thus of great importance for applications to have practical running times.

\IEEEpubidadjcol

\section{Methods}
\subsection{Programs}
Cellular automata can be programmed in many ways. Two of the major determinants of running time are how the cell states are represented and how they are updated according to a predefined rule or update function. Elementary cellular automata are one dimensional grids, and their cells are either off or on, thus 0 or 1. In one update iteration, the cell states are updated according to the update function, which results in a new one dimensional grid. In all programs and experiments of the present study, an integer cell state and the following update function are used.

\begin{lstlisting}
int update(int l, int c, int r)
{
   return (l + r - l * r) * (rand() % 2);
}
\end{lstlisting}

It is rule 250 in the algebraic definition \cite{Wolfram.NKS.2002}. \texttt{l}, \texttt{c}, and \texttt{r} are variable names for the cell states of left, centre, and right cells, respectively. To avoid compiler optimisation of this simple update rule, it is randomly multiplied by either one or zero.

Usually, cell states are represented by a Boolean type in simple cellular automata. However, because C++ vectors are tested here and \texttt{std::vector<bool>} is a specialisation of \texttt{std::vector} for the Boolean type, which is possibly optimised for space-efficiency \cite{CPPREF.vecbool,Stroustrup.Language.2013}, integer is used instead \cite{GFG.problemvecbool}. \texttt{std::bitset} and \texttt{std::vector<bool>} hold and access bits through proxy elements allowing a compact storage for a sequence of bits but might result in longer running times because of this indirect and more elaborate access \cite{Stroustrup.Language.2013}. \texttt{std::vector<bool>} is thus not comparable to an array but to a bitset with dynamic size \cite{Stroustrup.Language.2013}. Using any other type with vector is comparable to a dynamic array because both provide direct and efficient access to elements and these elements are stored continuously in memory \cite{Stroustrup.Language.2013}.

Since cellular automata divide space into cells and the state update depends on neighbouring cells, it is convenient to use arrays or a similar data structure to represent the space of a cellular automaton. The simplest way to represent a cell state is to use a primitive data type, however, one might use structs or objects to represent multiple states per cell. The questions are therefore how expensive it is to use structs or objects as well as how to loop over the cell states in the data structure to update them in each iteration. In the following, different versions in C and C++ of how to answer these questions are presented.

In all the programs, the cellular automaton size (grid size) is given as an integer, the memory of the grid is allocated before the update and is not part of the running time measurement, the initial grid is initialised with random values, and a for-loop is used to update cell states by calling the update function.

\subsubsection{Program 1 in C (C Exp 1) -- array of integer type, looped over by index}
The most simple and intuitive way to program a cellular automaton in C is to use an array with elements of primitive data type integer and the array index to loop over it. Since the cellular automaton size is known and constant, and cells have fixed neighbouring cells in elementary cellular automata, arrays are a suitable data structure here. Since arrays are closely representing memory, it is also expected to be an efficient program.
\begin{lstlisting}
int grid_size = 1000001;
int grid_0[grid_size];
int grid_1[grid_size];
for(int i=1; i<grid_size-1; ++i)
{
   grid_1[i] = update( grid_0[i-1], 
      grid_0[i], grid_0[i+1] );
}
\end{lstlisting}

\subsubsection{Program 2 in C (C Exp 2) -- array of integer type, looped over by pointer}
In C, it is also possible to access arrays with pointers instead of array subscripting \cite{RitchieKernighan.C.1988}. Instead of using the array index to loop over the array, the second program in C uses pointers to achieve the same. After defining the integer grids, pointers to the start and end cells are created. In the loop, the pointers to the grids are incremented moving them from one memory location to the next.
\begin{lstlisting}
int grid_size = 1000001;
int grid_0[grid_size];
int grid_1[grid_size];
int *p1_grid_0 = &grid_0[1];
int *p2_grid_0 = &grid_0[0]+grid_size-1;
int *p_grid_1 = &grid_1[1];
for( ; p1_grid_0 < p2_grid_0; ++p1_grid_0, 
   ++p_grid_1)
{
   *p_grid_1 = update( *(p1_grid_0-1),
      *p1_grid_0, *(p1_grid_0+1) );
}
\end{lstlisting}

\subsubsection{Program 3 in C (C Exp 3) -- array of integer structures, looped over by index}
In the present study, we are also looking at multi-state cellular automata, which means at data structures that can hold more than one state. This can be achieved with structures in C \cite{RitchieKernighan.C.1988}. It is programmed here by first defining a structure called cell with an integer member called state. Second, the grid array is an array of cell structures. And third, the structures in the array are accessed using the structure member operator \texttt{structure.member} in the for-loop.
\begin{lstlisting}
int grid_size = 1000001;
struct cell {
   int state;
};
struct cell grid_0[grid_size];
struct cell grid_1[grid_size];
for(int i=1; i<grid_size-1; ++i)
{
   grid_1[i].state = update(
      grid_0[i-1].state, grid_0[i].state,
         grid_0[i+1].state );
}
\end{lstlisting}

\subsubsection{Programs 1 and 2 in C++ (C++ Exp 1 and 2) -- array of integer type, looped over by index and pointer}
Programs 1 and 2 in C++ are the same as programs 1 and 2 in C but compiled with a C++ compiler. 

\subsubsection{Program 3 in C++ -- vector of integer type, looped over by index}
The C++ standard library provides data containers such as vector, which can hold a sequence of elements of a given type like an array, but it also allows extending the vector, it knows its size, and it can be checked against attempts to access out-of-range elements \cite{Stroustrup.Language.2013,Stroustrup.Principles.2014}. These features make vector the recommended data container in C++ and is thus used in the following program. The vector with elements of integer type and of predefined size is pre-allocated before the for-loop in which the elements are accessed by index.
\begin{lstlisting}
int grid_size = 1000001;
vector<int> grid_0(grid_size);
vector<int> grid_1(grid_size);
for(int i=1; i<grid_size-1; ++i)
{
   grid_1[i] = update( grid_0[i-1], 
      grid_0[i], grid_0[i+1] );
}
\end{lstlisting}

\subsubsection{Program 4 in C++ (C++ Exp 4) -- vector of integer type, looped over by iterator}
Instead of using an index, vector provides iterators that can be used to loop over it and some useful related functions \cite{Stroustrup.Language.2013,Stroustrup.Principles.2014}. In program 4 in C++, iterators of type integer are first created. The iterator \texttt{itr\_grid0} points to the first element of the vector and the iterator \texttt{last} points to the last element of the vector. For the sake of simplicity, boundaries of the cellular automata are not treated and thus \texttt{itr\_grid0} is incremented and \texttt{last} is decremented. In the for-loop, the initialisation is empty because it is done before the loop already, \texttt{itr\_grid0} and \texttt{last} are compared in the condition, and \texttt{itr\_grid0} and \texttt{itr\_grid1} are incremented in the update. Within the for-loop, the update function is called with values for the initial grid vector by dereferencing the iterator with \texttt{*itr\_grid0}. The return value of the update function is then assigned to the element at the respective position in \texttt{grid\_1} via iterator \texttt{*itr\_grid1}.
\begin{lstlisting}
int grid_size = 1000001;
vector<int> grid_0(grid_size);
vector<int> grid_1(grid_size);
vector<int>::iterator itr_grid0 =
   grid_0.begin() + 1;
vector<int>::iterator last =
   grid_0.end() - 1;
for( ; itr_grid0 != last;
   ++itr_grid0, ++itr_grid1)
{
   *itr_grid1 = update( *(itr_grid0-1),
      *itr_grid0, *(itr_grid0+1) );
}
\end{lstlisting}

\subsubsection{Program 5 in C++ (C++ Exp 5) -- array of integer objects, looped over by index}
C++ is object-oriented and thus provides classes to program multi-state cellular automata. Classes in C++ are like structures in C but in addition to member data, classes allow member functions. In program 5 in C++, a class \texttt{Cell} is first defined and subsequently used as a user-defined type for the grids and accessed by index and object member operator \texttt{object.member} in the for-loop.
\begin{lstlisting}
int grid_size = 1000001;
class Cell
{
   public:
      int state;
};
Cell grid_0[grid_size];
Cell grid_1[grid_size];
for(int i=1; i<grid_size-1; ++i)
{
   grid_1[i].state = update(
      grid_0[i-1].state, grid_0[i].state, 
         grid_0[i+1].state);
}
\end{lstlisting}

\subsection{Experiments}
Each of the C and C++ programs were run in experiments with corresponding numbering on a Mac Pro (2019, 2.5 GHz 28-Core Intel Xeon W, 1.5 TB 2933 MHz DDR4), each with the GNU gcc 14.2.0 and Apple clang 16.0.0 compilers. Compiler optimisation \texttt{-O2} was used. They were run for $10^4$ and $10^6$ cellular automaton grid sizes. Running time is measured using \texttt{clock()} of \texttt{time.h} because it can be used in C and C++. Only the for-loop is considered in the running time measurement.

\section{Results}
\begin{figure*}[t]
\centerline{\includegraphics[width=\textwidth]{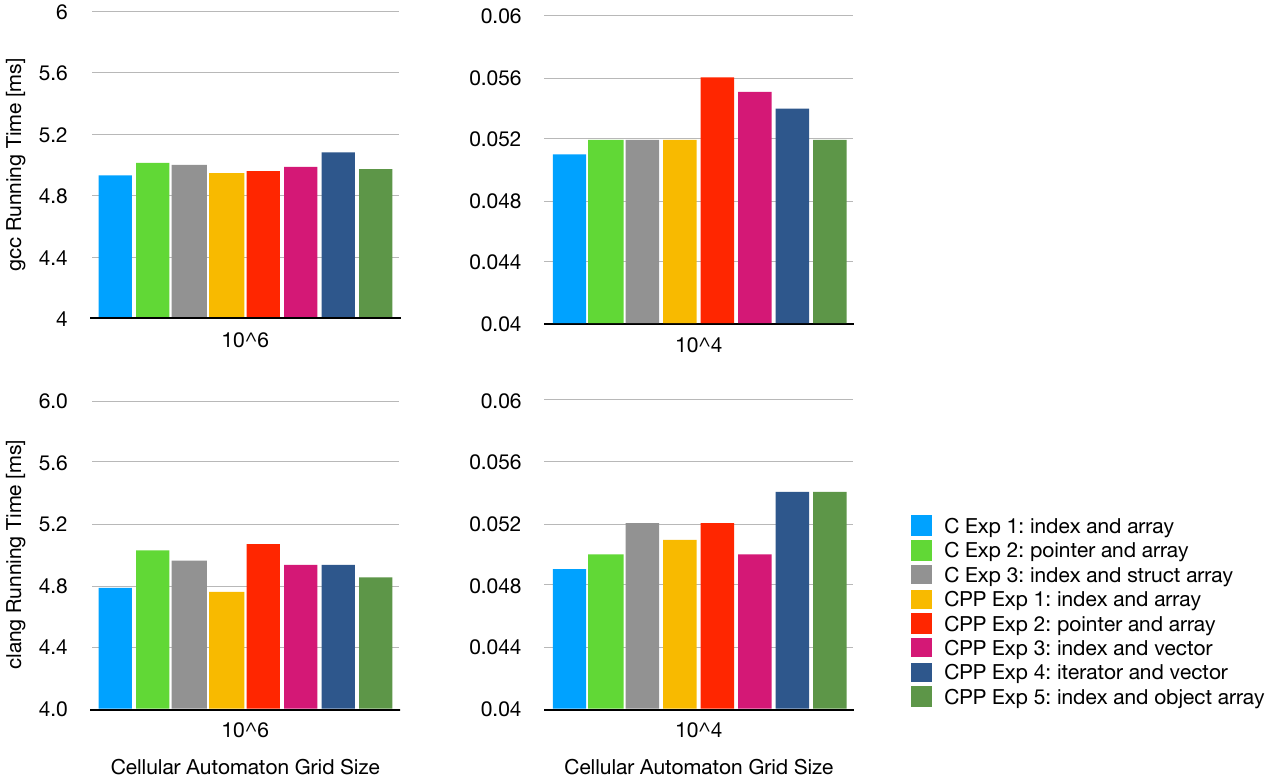}}
\caption{Running times in milliseconds for different cellular automata programs in C and C++. The top row shows the results for the GNU gcc compiler and the bottom row for the Apple clang compiler. The left column shows the results for a cellular automaton grid size of $10^6$ and the right column of grid size $10^4$.}
\label{fig1}
\end{figure*}

Running times of the eight experiments for the Apple clang and GNU gcc compilers are shown in fig.~\ref{fig1}. The top row shows the outcome for the GNU gcc compiler and the bottom row for the Apple clang compiler. The first and second columns show the results for grid size $10^6$ and $10^4$, respectively.

The measured running times show only minor differences between the tested programs. The choice of language (C versus C++), data structure (array versus vector), data type (integer versus integer struct/class), and compiler (gcc versus clang) did not affect the running time. In addition, running times scale linearly with grid size for both compilers as shown by comparing the first and second column in fig.~\ref{fig1}.

It is interesting to note that the Apple clang compiler was able to optimise simple update functions, e.g. rule 250 without random multiplication by zero or one. It was also able to do so if the cellular automata rule was represented by conditionals. This optimisation showed three orders of magnitudes faster running times for the tested programs except for the vector program with iterators, which showed two orders of magnitudes decrease in running time.

\section{Discussion}
The present study explores running times of different ways to program cellular automata in C and C++. Specifically, looping through arrays by different means, the effect of structures and objects, and the choice of data structure (array versus vector in C++) and compiler (GNU gcc versus Apple clang) were studied. The tested programs show only negligible differences in running times. Only the compiler choice has an effect in the case of simple update functions.

In practice, it is highly recommended to use vector instead of array in C++ because they are safer, dynamic, more convenient, and are nevertheless efficient \cite{Stroustrup.Language.2013,Stroustrup.Principles.2014}. Most developers are indeed using vector as their default data container. The present study supports this practice by showing that vectors are indeed as performant as arrays if their memory is pre-allocated. In addition, the study hints to access vectors by index and not iterators to exploit compiler optimisation of simple update functions. However, this needs further investigation and confirmation.

Regarding multi-state cellular automata or the use of structure and object, the experiments show that both are implemented very efficiently as they have hardly any effect on running time. This is an encouraging result to design programs using structures in C and classes in C++. In contrast, in C\#, using objects does significantly influence running time \cite{Christen.SMC.2022}.

There are limitations to this study, especially regarding its scope. It would be most interesting to investigate compiler optimisation, parallel programming, and Apple's Metal shader. Future studies might study these aspects of programming cellular automata. In this study, I wanted to start with some fundamental aspects, which might be useful to build up in future studies.

In conclusion, it is recommended to use vector accessed by index in C++ because it is safe and performant. If one is interested in multi-state cellular automata, objects can be used without loss of that performance. Future studies might investigate Apple's Metal shader or compiler optimisation, especially with respect to the update function.

\section*{Acknowledgment}
Seth Hillbrand from the open source KiCad EDA project is gratefully acknowledged for providing feedback on an earlier version of the manuscript. The Santa Fe Institute is gratefully acknowledged for hosting the author as a visitor in Autumn 2023.

\bibliographystyle{IEEEtran}
\bibliography{bibliography}

\begin{thebibliography}{10}
\providecommand{\url}[1]{#1}
\csname url@samestyle\endcsname
\providecommand{\newblock}{\relax}
\providecommand{\bibinfo}[2]{#2}
\providecommand{\BIBentrySTDinterwordspacing}{\spaceskip=0pt\relax}
\providecommand{\BIBentryALTinterwordstretchfactor}{4}
\providecommand{\BIBentryALTinterwordspacing}{\spaceskip=\fontdimen2\font plus
\BIBentryALTinterwordstretchfactor\fontdimen3\font minus
  \fontdimen4\font\relax}
\providecommand{\BIBforeignlanguage}[2]{{%
\expandafter\ifx\csname l@#1\endcsname\relax
\typeout{** WARNING: IEEEtran.bst: No hyphenation pattern has been}%
\typeout{** loaded for the language `#1'. Using the pattern for}%
\typeout{** the default language instead.}%
\else
\language=\csname l@#1\endcsname
\fi
#2}}
\providecommand{\BIBdecl}{\relax}
\BIBdecl

\bibitem{Wolfram.Nature.1984}
S.~Wolfram, ``Cellular automata as models of complexity,'' \emph{Nature}, vol.
  311, pp. 419--424, 1984.

\bibitem{Wolfram.NKS.2002}
------, \emph{{A New Kind of Science}}.\hskip 1em plus 0.5em minus 0.4em\relax
  Champaign: Wolfram Media, 2002.

\bibitem{Hubert.IEEESpectrum.2024}
B.~Hubert, ``Why bloat is still software's biggest vulnerability: A 2024 plea
  for lean software,'' \emph{IEEE Spectrum}, vol.~61, no.~4, 2024.

\bibitem{Wirth.Computer.1995}
N.~Wirth, ``{A Plea for Lean Software},'' \emph{Computer}, vol.~28, no.~2, pp.
  64--68, 1995.

\bibitem{Wirth.IEEEAHC.2008}
------, ``{A Brief History of Software Engineering},'' \emph{IEEE Annals of the
  History of Computing}, vol.~30, no.~3, pp. 32--39, 2008.

\bibitem{CPPREF.vecbool}
``{C++ reference -- C++, Containers library: std::vector\textless
  bool\textgreater},''
  \url{https://en.cppreference.com/w/cpp/container/vector\_bool} [Accessed: 16
  October 2024].

\bibitem{Stroustrup.Language.2013}
B.~Stroustrup, \emph{{The C++ Programming Language}}, 4th~ed.\hskip 1em plus
  0.5em minus 0.4em\relax Pearson, 2013.

\bibitem{GFG.problemvecbool}
``{GeeksforGeeks -- Problem With std::vector\textless bool\textgreater in
  C++},''
  \url{https://www.geeksforgeeks.org/problem-with-std-vector-bool-in-cpp/}
  [Accessed: 16 October 2024].

\bibitem{RitchieKernighan.C.1988}
D.~M. Ritchie and B.~W. Kernighan, \emph{{The C Programming Language}},
  2nd~ed.\hskip 1em plus 0.5em minus 0.4em\relax Pearson, 1988.

\bibitem{Stroustrup.Principles.2014}
B.~Stroustrup, \emph{{Programming: Principles and Practice Using C++}},
  2nd~ed.\hskip 1em plus 0.5em minus 0.4em\relax Pearson, 2014.

\bibitem{Christen.SMC.2022}
P.~Christen, ``{Programming Data Structures for Large-Scale Desktop Simulations
  of Complex Systems},'' in \emph{2022 IEEE International Conference on
  Systems, Man, and Cybernetics (SMC)}, 2022, pp. 3000--3005.

\end{thebibliography}

\end{document}